\documentclass[prb,aps,twocolumn,epsfig,eqsecnum]{revtex4}

\usepackage{epsfig}
\usepackage{wasysym}
\usepackage{bm}
\usepackage{pstricks}

\newcommand{\beq}{\begin{equation}}
\newcommand{\eeq}{\end{equation}}
\newcommand{\bea}{\begin{eqnarray}}
\newcommand{\eea}{\end{eqnarray}}

\newcommand{\etal}{{\em et al.}}

\def\tit#1#2#3#4#5{{#1}{\bf #2}, #3 (#4)}
\def\jmp{J.\ Math.\ Phys.\ }

\def\prl{Phys.\ Rev.\ Lett.\ }
\def\pr{Phys.\ Rev.\ }
\def\prb{Phys.\ Rev.\ B\ }

\def\jpco{J.\ Phys.\ Cond.\ Mat.\ }

\def\jpa{J.\ Phys.\ A\ }

\def\jpsj{J.\ Phys.\ Soc.\ Jpn.\ }
\def\sci{Science\ }
\def\natu{Nature\ }

\def\jsp{J.\ Stat.\ Phys.\ }

\def\vecb#1{{\mathbf {#1}}}

\begin{document}


\title{Magnetization process of spin ice in a [111] magnetic field}

\author{S. V. Isakov,$^1$ K. S. Raman,$^2$ R. Moessner$^3$ 
and S. L. Sondhi$^2$}

\affiliation{$^1$Department of Physics, AlbaNova, Stockholm University,
SE-10691 Stockholm, Sweden}

\affiliation{$^2$Department of Physics, Princeton University,
Princeton, NJ 08544, USA}

\affiliation{$^3$Laboratoire de Physique Th\'eorique de l'Ecole Normale
Sup\'erieure, CNRS-UMR8549, Paris, France}

\date{\today}

\begin{abstract}
  Spin ice in a magnetic field in the [111] direction displays two
  magnetization plateaux, one at saturation and an intermediate one with
  finite entropy. We study the crossovers between the different regimes from a
  point of view of (entropically) interacting defects. We develop an
  analytical theory for the nearest-neighbor spin ice model, which covers most
  of the magnetization curve. We find that the entropy is non-monotonic,
  exhibiting a giant spike between the two plateaux. This regime is
  described by a monomer-dimer model with tunable fugacities. At low fields,
  we develop an RG treatment for the extended string defects, and we compare
  our results to extensive Monte Carlo simulations. We address the
  implications of our results for cooling by adiabatic (de)magnetization.
\end{abstract}


\maketitle

\section{Introduction}
Recent experiments on the spin ice
compounds\cite{harspinice,ramspinice} Ho$_2$Ti$_2$O$_7$ and
Dy$_2$Ti$_2$O$_7$ have uncovered an intriguing set of phenomena when
unicrystalline samples are placed in an external magnetic field in the
[111] direction.\cite{matsuice,hiroiice,higaice,fukazawaice} For a review
on spin ice, see Ref.~\onlinecite{ginbrarev}.

The discovery of a plateau in the magnetization below saturation, first
predicted theoretically\cite{liquidgas} and explored in Monte Carlo
simulations,\cite{liquidgas,dipolerahul} has been particularly remarkable as
it was found to retain a fraction of the zero-field spin ice
entropy.\cite{hiroiice,ogata,msspinice} In this regime, the system is well
described by a two-dimensional Ising model on a kagome lattice in a
longitudinal field, which is in turn equivalent to a hexagonal lattice dimer
model.\cite{ogata,msspinice,isingquant}.

Recently, two of the present authors have studied the thermodynamics and
correlations of the [111] plateau.\cite{msspinice} This work has led to the
identification of mechanisms which terminate the plateau. At the high-field
end, the termination occurs via the proliferation of monomer defects in the
underlying dimer model.  At low fields, a more exotic extended string defect
restores three dimensionality. The asymptotic density of both kinds of defects
was estimated in Ref.~\onlinecite{msspinice}.

\begin{figure}[ht]
{
\centerline{\includegraphics[angle=0, width=2.8in]{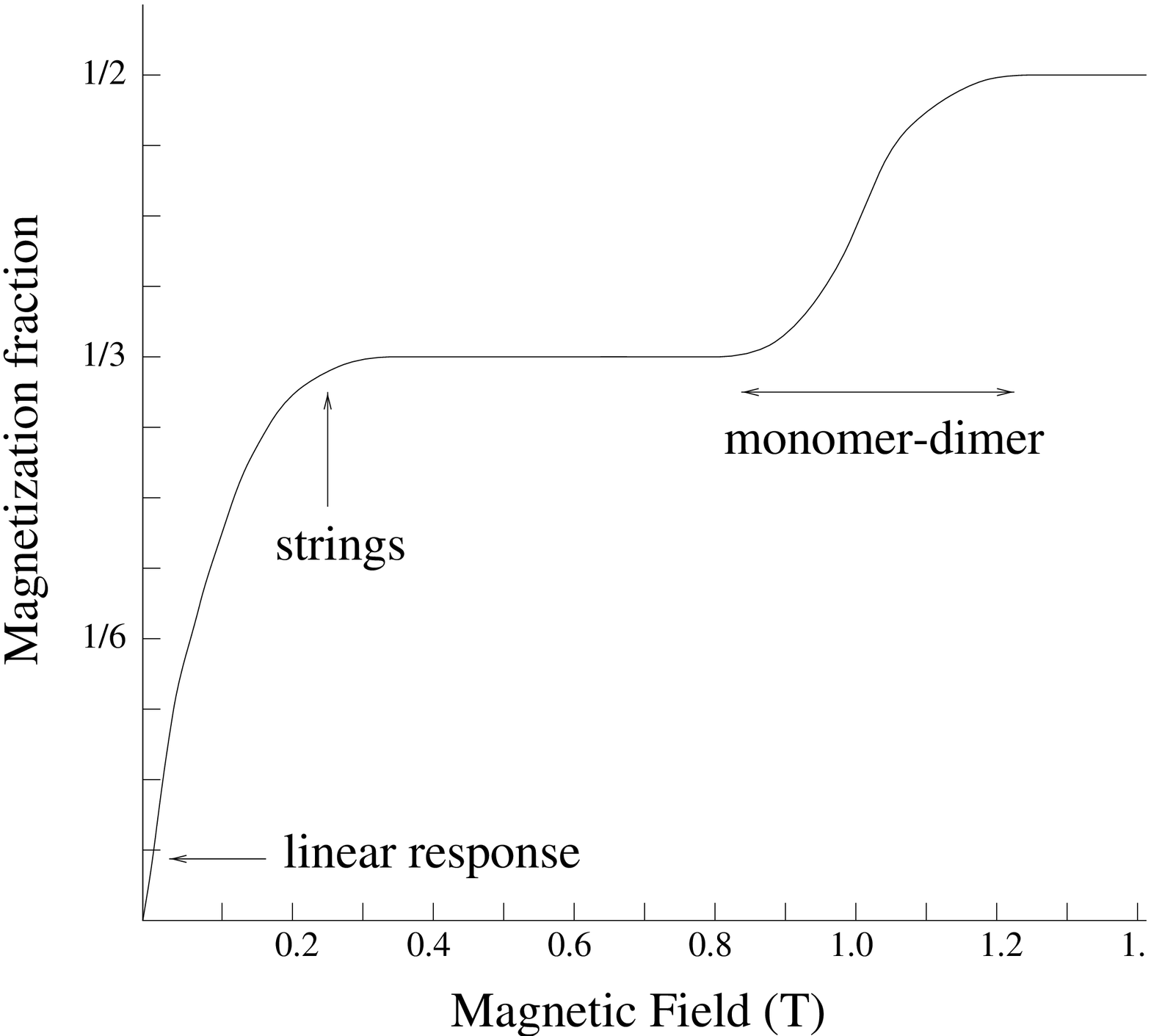}}
\centerline{\includegraphics[angle=270, width=2.8in]{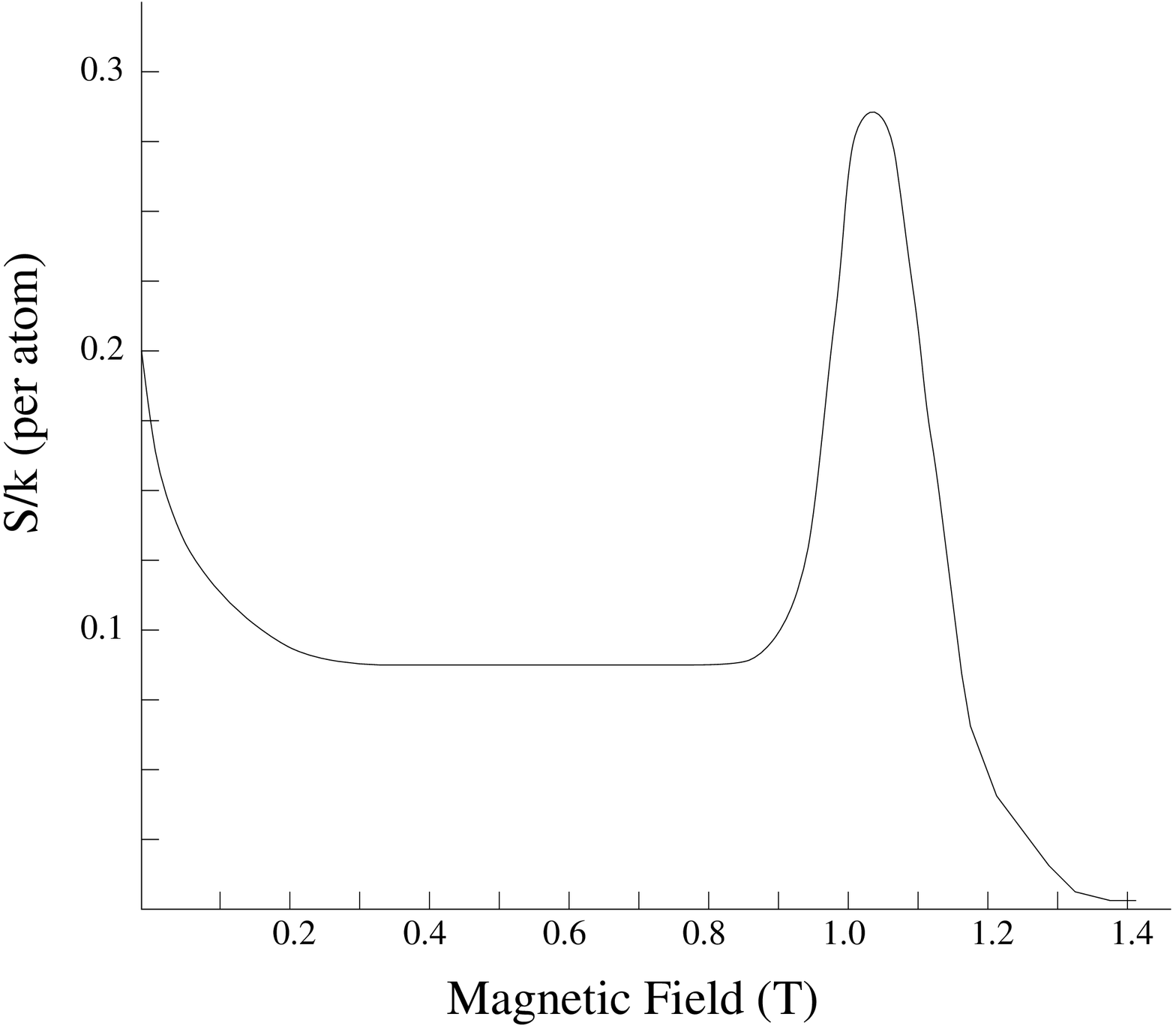}}
\caption{Properties of spin-ice as the [111] magnetic field is varied.  
These curves are for illustration and do not show actual numerical or 
experimental data.  We have indicated the regions where various analytic
approaches discussed in the text apply. 
}
\label{fig:cartoon}}
\end{figure}

In this paper, we consider in detail the full magnetization curve from
zero-field to saturation. A brief synopsis of the exotic thermodynamic
properties of spin ice is in a $[111]$ field is sketched in
Fig.~\ref{fig:cartoon}. The aim of this paper is to identify the different
regimes of the magnetization curves, to provide analytical theories for them,
and to test them against Monte Carlo simulations, and finally against
experiment.

Near zero field, we use the accurate self-consistent Hartree
approximation\cite{imslargen} to provide an analytical approximation for 
the linear response regime.  At the low field end of the plateau, we
develop mean field and renormalization group treatments for the
extended string defects, which we use to analyze the in-plane and
out-of-plane correlations.  We compare these with Monte Carlo
simulations using an efficient cluster algorithm, which allows us to
obtain accurate data from the linear response regime to the beginning
of the [111] plateau.  We find that the mean field treatment is
accurate at the lowest fields, where the string density would be
relatively high.  The renormalization group treatment compares well
with simulation in the dilute string limit.  At even higher fields,
the plateau is approached and the suppression of the entropic
activation of strings becomes apparent as a finite-size effect.

\begin{figure}[ht]
{\begin{center}
\includegraphics[width=3.2in]{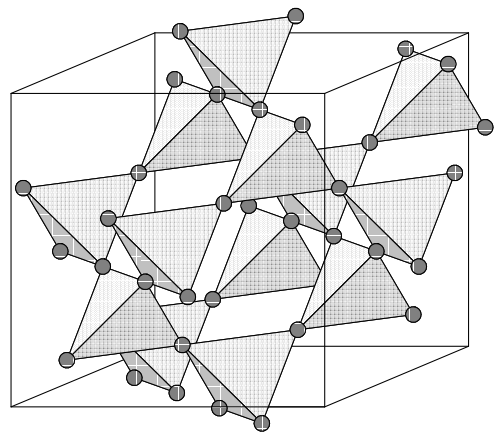}
\vspace{-4cm}
\caption{The pyrochlore lattice of corner-sharing tetrahedra.}
\label{fig:pyro}
\end{center}}
\end{figure}

At the high-field termination of the plateau, we observe a giant peak in the
entropy, which even exceeds the zero field Pauling value, despite the fact
that a quarter of all spins are pinned. We model this phenomenon by a
monomer-dimer model on the honeycomb lattice with varying fugacities. (At the
point where the all fugacities equal 1, this model turns out to be one of
`hard bow-ties' on the kagome lattice.) We analyze this model within a Bethe
approximation and also by using results from a high-order series
expansion.\cite{naglecorr}

We show that the entropy peak is due to the crossing of an extensive number of
energy levels which have macroscopic entropies. Comparing this theory with
Monte Carlo simulations of the appropriate monomer-dimer model, we find that
the simple Bethe approximation is accurate for moderate to large monomer
densities. 

We point out that this theory predicts to a crossing
point in the plots of magnetisation versus field at different
temperatures. In addition, there is a further crossing point at lower fields,
where the corrections to the magnetisation due to monomer and string defects
almost cancel one another.

We then address the connection of these results to experiment, in particular
pointing out the presence of (at least a vestige) of the entropy peak in
existing experimental data.

We then discuss the implications of the entropy peak for magnetocaloric
manipulations. In particular, we argue that it arises in a more general set of
models. It can, in principle, be used to effect cooling in a field, both by
adiabatic demagnetization, and by adiabatic magnetization.
Finally, we close with some concluding remarks.

\section{Model and notation}

\begin{figure}[ht]
{\begin{center}
\includegraphics[width=1.6in]{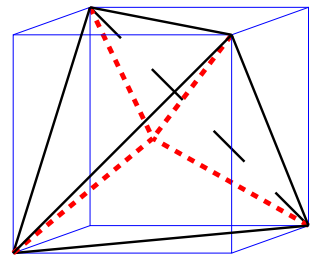}
\caption{A single tetrahedron inscribed in a cube. The easy axes of the
pyrochlore lattice (or $\langle 111 \rangle$ axes),
${\vecb{\hat{d}}}_{\kappa}$, are indicated by the short-dashed lines.
}
\label{fig:tetrahedron}
\end{center}}
\end{figure}

A general model of spin ice includes the single-ion anisotropy, the exchange
interaction, and the dipolar interaction. In this work we use a simplified
model\cite{harspinice} in which the long-range dipolar interaction is
truncated beyond the nearest-neighbor spins. While the exchange interaction in
spin ice compounds is antiferromagnetic, the effective interaction
(exchange plus nearest-neighbor dipolar) is ferromagnetic. The Hamiltonian
for unit-length spins $\vecb{S}_{i}$ may be written as
\bea
  H &=& -J_{\rm eff}' \sum_{<ij>} \vecb{S}_{i} \cdot \vecb{S}_{j}
    + E \sum_{i} \left({\vecb{\hat{d}}}_{\kappa(i)}
          \cdot \vecb{S}_{i} \right)^{2} \nonumber \\
    &+& g \mu_{B} J \sum_{i} \vecb{B} \cdot \vecb{S}_{i},
\label{heff1}
\eea
where $J_{\rm eff}'$ is an effective exchange coupling.  The second term
is the easy axis anisotropy of strength $E < 0$, $|E|\agt 50K$ , which
is much larger than the exchange and dipolar interaction strengths.  The
unit vectors ${\vecb{\hat{d}}}_{\kappa(i)}$ are the local easy axes of the
pyrochlore lattice (see Fig.~\ref{fig:pyro} and
Fig.~\ref{fig:tetrahedron}). The third term is the
interaction with a magnetic field of strength
$B$, $g \mu_{B} J$ being the magnetic moment of the spins. Both experiment
and theory indicate that this simplified model is a good
description of spin ice at moderate temperatures.

In our analysis, we take the single ion anisotropy to be infinite so the spins
are constrained to lie along their local easy axes.  In this limit, it is 
convenient to describe the system by the Ising
pseudospins $\sigma_{i}$, where 
$\vecb{S}_{i}=\sigma_i {\vecb{\hat{d}}}_{\kappa(i)}$.
The pseudospin $\sigma_{i}=$+1(-1) if the physical spin points into (out of)
its associated up-pointing tetrahedron.  We may write an effective 
Hamiltonian for the pseudospins:
\bea
  H = J_{\rm eff} \sum_{<ij>} \sigma_{i} \sigma_{j}
    -g \mu_{B} J \sum_{i} \vecb{B} \cdot \vecb{\hat{d}}_{\kappa(i)} \sigma_{i},
\label{heff2}
\eea
where $J_{\rm eff} = J_{\rm eff}'/3 > 0$. 

\section{The low field regime}
\noindent
At zero magnetic field and zero temperature, the ferromagnetic interaction
gives rise to an ``ice rule'' constraint: the pseudospins on each tetrahedron
must sum to zero, $\left| \sum_{\kappa} \sigma_{\kappa} \right| = 0$.  In
terms of the physical spins, on each tetrahedron two will point inwards
(towards the center) and two will point outwards (away from the center).  The
set of configurations satisfying the ice rule comprises the zero-field spin
ice ground state manifold.  At low magnetic fields (and low temperatures), the
system will continue to obey the ice rule, though the magnetic field will
favor certain states among those in the zero-field ground state manifold.

\begin{figure}[ht]
{
\centerline{\includegraphics[angle=0, width=3.2in]{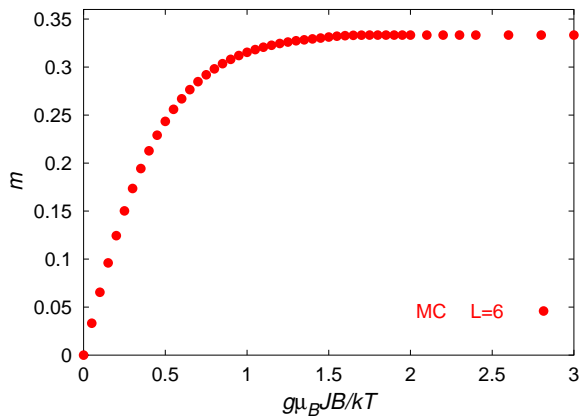}}
\caption{The magnetization from Monte Carlo simulations.}
\label{fig:icemag}}
\end{figure}

We have performed extensive Monte Carlo simulations of the low field regime,
from zero-field up till the low field plateau termination, using a loop
algorithm, which is discussed in Appendix A.  Our algorithm probes only spin
ice ground states (two spins in and two out on each tetrahedron) and is
thus
applicable at low temperatures $T\ll J_{\rm eff}$ and low magnetic fields,
where the density of monomer defects, which are responsible for the high field
plateau termination, is low.  The simulation is written in terms of a
pyrochlore lattice with the conventional 16 site cubic unit cell (which
contains four tetrahedra of each kind). The simulations have been done for
systems with $16$, $128$, $432$, $1024$, $2000$, $3456$, $5488$, $8192$, and
$16000$ sites. For a system with $16000$ sites, we perform $2.5\times10^6$
loop flips for equilibration and $5\times10^7$ for averaging. For other system
sizes, we we perform $1\times10^7$ loop flips for equilibration and
$2\times10^8$ for averaging. The simulated magnetization as a
function of the magnetic field strength is shown in Fig.~\ref{fig:icemag}. The
magnetization attains the plateau value at fields much larger than the
temperature.

\subsection{The linear response regime}
\noindent
We may calculate the ground state entropy of spin ice at zero
field by numerically integrating the first law of thermodynamics
\beq
        dS=\frac{dU}{T}+\frac{m}{T}dh.
\eeq
Noting that the magnetization is constant and equal to $-g\mu_{B}J/3$ per
spin on the plateau and is zero at zero field, 
and that the value of the entropy on the plateau
is $S/k_{B}=0.080765$\cite{ogata, msspinice}, we obtain for the entropy of
spin ice, $S/k_{B}=0.2051\pm0.0001$. Our value is very close to Pauling's 
estimate $S/k_{B}=0.202733$ and is consistent with the most accurate 
current theoretical estimate $S/k_{B}=0.20501\pm0.00005$.\cite{nagleent}

At zero field, we use the self-consistent Hartree approximation, which is
known to give a quantitatively accurate approximation to the 
ground state correlations
of spin ice.\cite{imslargen} This gives $\chi=2(g\mu_{B}J)^2/3k_{B}T$ for
spin ice.  This compares well with our Monte Carlo result,
$\chi=(0.66735\pm0.0003)(g\mu_{B}J)^2/k_{B}T$ for a system with 16000 sites.

\subsection{String defects and their interactions}
\subsubsection{General description}
Figure \ref{fig:pyro} presents the underlying pyrochlore lattice of spin ice
and fig. \ref{fig:tetrahedron} shows the [111] direction.  It is convenient
to visualize the pyrochlore lattice as a stack of alternating kagome
and triangular planes, the [111] direction being the direction in
which the planes are stacked.  Each spin lies on a corner shared by an
up-pointing and down-pointing tetrahedron.

If the [111] magnetic field is large enough, the spins in the triangular 
planes align with the field; the kagome planes decouple from 
one another; and the system is well described by a two-dimensional 
model.  This describes spin ice on the plateau.  At fields slightly lower
than the plateau, excitations called string defects,\cite{msspinice} restore 
three-dimensionality and are responsible for the low field termination of the 
plateau. 

To describe these defects, we consider the entropic benefit of relaxing the
condition that the triangular planes are polarized. 
Suppose we flip a spin in some triangular layer.  Then, by the ice rule 
constraint, we must also flip a spin in each of the two neighboring kagome 
layers (on the two tetrahedra that are sharing the first flipped spin).  
Flipping these kagome spins requires flipping spins in each of the two 
neighboring triangular layers, which requires flipping spins in the two 
next-nearest kagome layers and so on.  The resulting ``string defect'' is
an excitation that extends through the system.  The energy cost, per 
kagome-triangle bilayer, of creating the string is $E_s=8g\mu_{B}JB/3$.
To estimate the entropy, we note that creating a string actually involves 
creating a pair of defects in each kagome plane.  A ``positive'' defect 
connects the kagome plane to the kagome plane directly above it via a 
flipped spin in the intermediate triangular plane.   Similarly, a ``negative''
defect connects the kagome plane to the kagome plane directly below it.  
These two defects may be separated by flipping pairs of spins pointing in 
different directions on neighboring triangles of 
the kagome plane.  The entropy in the kagome plane depends on this 
separation, which is the basis for the interaction between defects discussed
below.  Ignoring this correction, the positive defect may be placed anywhere
in the plane (which fixes the position of the negative defect in the layer
above).  This implies that the entropy per bilayer is $S\sim\ln A$, 
where $A$ is the area of a layer.  This shows that for a given magnetic 
field, string defects are favored in a sufficiently large system.  For a
given system size, strings are favored at sufficiently low magnetic 
fields.  

\subsubsection{Interactions}
For magnetic fields in the plateau region, the triangular spins are fixed while
each kagome plane contains two up pseudospins ($\sigma=1$) and one down
pseudospin ($\sigma=-1$). This Ising model on the kagome lattice may be mapped
onto the dimer model on the hexagonal 
lattice\cite{ogata,msspinice,isingquant}, where
a down pseudospin corresponds to a dimer on the hexagonal lattice.  In this 
language, a string defect appears as a pair of oppositely charged monomers.

As discussed in Ref.~\onlinecite{msspinice}, a monomer-dimer
covering may be described by assigning a height variable $h_i$ to each site
$i$ of the triangular lattice dual to the hexagonal lattice on which the
dimers lie.  The heights are assigned as follows.  Moving from a site 
to a nearest neighbor site by moving clockwise around an up- (down-) triangle
will increase (decrease) the height by +2 (-2) if a dimer is crossed.  If
a dimer is not crossed, then the height will decrease (increase) by -1 (+1).
According to these rules, traversing a closed loop in the dual lattice will 
result in a height difference of +3 (-3) if a positive (negative) monomer is 
enclosed and 0 otherwise.  We note that the overall sign of the height 
assignments is a matter of convention and we may as well have chosen the $h_i$
so that traversing a closed loop containing a positive (negative) monomer 
gave a height difference of -3.  

In a coarse-grained description, the $h_i$ are replaced by a real, continuum
field $h(\vec{r})$ and as discussed in Ref.~\onlinecite{msspinice}, 
the entropy associated with a height field $h(\vec{r})$ is given to lowest 
order ingredients by:
\beq
S=\int d^2r \frac{K}{2}|\nabla h|^2
\eeq
where $K=\frac{\pi}{9}$ for the honeycomb lattice.\cite{coarse}  
The height field has the
property:
\beq
\oint_C \nabla h\cdot d\vec{r}=3\int_S d^2r \sigma(\vec{r})
\eeq
where $\sigma(\vec{r})$ is the monomer charge density and $S$ is the 
region enclosed
by the loop C.  We may proceed by analogy with the 2d XY model\cite{piers} and
divide $h$ into ``dimer'' (spin-wave) and ``monomer'' (vortex) contributions.
A standard calculation\cite{kumar} gives the entropy of the monomer piece:
\begin{eqnarray}
S_m&=&\frac{9K}{4\pi}\int\int d^2r d^2r' \sigma(\vec{r})\sigma(\vec{r'})\Bigl(-\ln\frac{|\vec{r}-\vec{r'}|}{\tau}\Bigr)\nonumber\\&=&\frac{1}{2}\int\int d^2r d^2r' \sigma(\vec{r})\sigma(\vec{r'})\Bigl(-\kappa\ln\frac{|\vec{r}-\vec{r'}|}{\tau}\Bigr)\nonumber\\
\end{eqnarray}
where $\kappa=1/2$ and $\tau$ is a hard-core radius comparable to the 
lattice spacing.  This shows that the entropic interaction between two
defects separated by distance $r$ is given by 
$p_1p_2 V(|\vec{r_1}-\vec{r_2}|)$ where $p_i$ is +1 (-1) for a positive
(negative) defect and $V(R)=-\kappa\ln(R/\tau)$.

\subsubsection{Mean field calculation}
If the number of defects is fairly large, we may expect the 
interaction to be sufficiently screened to justify the use of 
variational mean field theory.\cite{chailub}  We will investigate the 
in-plane and 
out-of-plane correlations for the defects. 

We consider a layered system of two-dimensional planes (indexed by the label 
$k$ which ranges from $-K$ to $K$) where each plane 
contains $N$ positive and $N$ negative defects (which we refer to as charges) 
that interact logarithmically.
The string constraint requires that each positive charge in layer $k$ is 
rigidly connected to a negative charge in the layer $k+1$.  We impose a 
periodic boundary condition to connect the positive charges in the $K$th 
layer to the negative charges in the $-K$th layer.

We formally impose the constraint by writing the ``Hamiltonian'' in terms of 
positive charges alone.  The planes are stacked in the
$z$-direction.  Let $x_i^k$ be the in-plane position of the $i$th 
positive charge in the $k$th layer.  In absence of external fields, the
entropy of a particular configuration of N defects is given by:
\begin{equation}
\label{mfh}
H=\sum_{k=-K}^K(\sum_{i \neq j}^N V(|x_i^k-x_j^k|) - \sum_{i,j}^N V(|x_i^k-x_j^{k+1}|) )
\end{equation}
Here $V(R)=-\kappa\ln (R/\tau)$, where $\tau$ is a hard-core radius defining 
the minimum separation between two charges and $\kappa=1/2$.  The first term
corresponds to the repulsion of positive charges within the same layer.  The 
absence of a factor of $\frac{1}{2}$ in front of this term is due to the string
constraint: bringing two positive charges in the same plane close together also
involves bringing together their negative partners in the plane above.  In 
terms of our positive charge formulation, this means the repulsion is twice 
as large.  The second term is the interlayer interaction.  Physically, a 
positive charge in layer $k$ has a negative partner in the layer $k+1$ which 
attracts the positive charges in layer $k+1$.  In terms of our positive 
charge formulation, charges repel charges in the same plane but attract 
charges in nearest neighbor planes.

We assume a variational mean field density of the form:
\begin{equation}
\rho(x_1^{-K},...,x_i^k,...,x_N^K)=\prod_{k=-K}^K\prod_{i=1}^N \frac{\rho^k(x_i^k)}{N}
\end{equation}
which asserts that all particles in a given layer $k$ have the same
probability density $\rho^k(x)/N$, but the 
density may vary from layer to layer.  We also need the normalizing condition:
\begin{equation}
\int_{A}d^2x \rho^k(x) = N
\end{equation}
This trial function implies a variational entropy functional:
\begin{eqnarray}
S_{\rho,N}&=&\sum_{k=-K}^K\Bigl(-\frac{1}{2}\int\int d^2x d^2x'(\rho^k(x)-\rho^{k+1}(x))\nonumber\\&\times&(\rho^k(x')-\rho^{k+1}(x'))V(|x-x'|)\nonumber\\&-&\int d^2x \rho^k(x)\ln(\frac{\rho^k(x)}{N})\Bigr)
\end{eqnarray}
This functional is maximized when the density is uniform 
$\rho^k(x)=\frac{N}{A}$ which gives $S_{\rho,N}=(2K+1)N\ln A$.  To investigate
the linear response of the system, we may apply a perturbing potential to the 
objects in the $k=0$ plane.  In particular, we consider the effect on the
density of placing a positive charge at the origin of the plane.  The details
are given in Ref.~\onlinecite{kumar} but we may quote the result:
\begin{eqnarray}
&&\delta\rho(\frac{x}{\xi_\parallel},k)=\frac{1}{4\pi^2\xi_\parallel^2}\nonumber\\&\times&\int\frac{ d^2s\Bigl[s^2(s^2+2)\Bigr]^{-1/2} e^{is\cdot(\frac{x}{\xi_\parallel})}}{\Bigl[1+s^2+2\sqrt{s^2(s^2+2)}\Bigr]\Bigl[1+s^2+\sqrt{s^2(s^2+2)}\Bigr]^{k-1}}\nonumber\\
\end{eqnarray}
where the in-plane length scale is given by $\xi_\parallel=\Bigl(\frac{A}{4\pi\kappa N}\Bigr)^{1/2}$.  We note first that this expression diverges at small $x$ for $k=0$,
which is not surprising because the assumption of a linear response would be not
be valid so close to the perturbing charge.  The expression would be valid at larger
$k$ and an interesting feature is that when $x=\xi_{\parallel}$, the decay in the 
$z$-direction does not depend on any physical parameters, i.e. there is no length scale
in the $z$ direction.  We will return to this point in the next section.

To connect with our physical problem, we note that at a given temperature, we will
have an expected value of defects which may be calculated from the partition
function:
\begin{equation}
\mathcal{Z}=e^{-\beta \mathcal{A}}=\sum_N \frac{y^{(2K+1)N}}{(N!)^{2K+1}}e^{S_N}
\end{equation}
where $S_N$ is the entropy of having $N$ defects and $y=e^{-E_s/k_BT}$ is the fugacity
of a positive defect ($y^{2K+1}$ is the fugacity of a ``string'').  In mean field,
we may replace $S_N$ by $S_{\rho,N}=(2K+1)N\ln A$.  From this, we may show\cite{kumar}
that $<N>\sim yA$ and using our earlier expression, we find that:
\begin{equation}
\xi_{\parallel,MF}^2\sim \exp(8g\mu_BJB/3k_BT)
\end{equation}   

\subsubsection{RG calculation}
When the gas of defects is fairly dilute, we may expect that the 
screening is not
effective enough to justify a mean field treatment.  In this section, 
we account for
fluctuations by making a real space renormalization group calculation using 
methods similar to the Kosterlitz treatment of 
the 2d Coulomb gas.\cite{piers,kt}

The dynamical objects described by Hamiltonian \ref{mfh} are dipoles of length
1.  We need to generalize this model in order to do an RG calculation.  The
generalization that we consider is allowing for dipoles of arbitrary length.
An ``$l$-dipole'' is an object where the negative charge lies directly $l$
planes above its positive partner.  While the original problem involved just
the coupling of nearest neighbor planes, our generalized model involves all
possible couplings.  Associated with each $l$-dipole is a fugacity $y_l/2\pi$
(the $2\pi$ is for convenience).  The grand partition function for the system
may be written as:
\begin{equation}
\label{grand}
\mathcal{Z}=\sum_{\{N_{k,l}\}}\Biggl[\prod_{k,l}\frac{(y_l/2\pi)^{N_{,l}}}{(N_{k,l})!}\Biggr]Z[\{N_{k,l}\}]
\end{equation}
where $N_{k,l}$ denotes the number of $l$-dipoles in layer $k$; $N_{,l}$ is the number of
$l$-dipoles in the system; and $N_k$ is the number of dipoles (of any length) that have
their positive ends in layer 
$k$.  The sum is over all particle number configurations $\{N_{k,l}\}$ that satisfy 
the charge neutrality constraint in each plane: $N_k=\sum_{l}N_{k-l,l}$.  The canonical 
partition function corresponding to a given dipole distribution $\{N_{k,l}\}$ is:
\begin{eqnarray}
\label{canon}
Z[\{N_{k,l}\}]&=&\int_{\Omega_{\tau}}\prod_{k,i}\Biggl(\frac{d^2x_{k,i}^{(1)}}{\tau^2}\frac{d^2x_{k,i}^{(2)}}{\tau^2}\delta\Bigl(\frac{x_{k,i}^{(1)}-x_{k,i}^{(2)}}{\tau}\Bigr)\Biggr)\nonumber\\&\times&\exp\Bigl[-H(\{N_{k,l}\})\Bigr]
\end{eqnarray}
$H(\{N_{k,l}\})$ is the Hamiltonian (actually an entropy) corresponding to 
the dipole distribution 
$\{N_{k,l}\}$.  The coordinate $x_{k,i}^{(1)}$ is the planar coordinate of the $i$th 
positive charge of layer $k$ and $x_{k,i}^{(2)}$ is the planar coordinate of its negative
partner which lives in layer $k+l(i)$, $l(i)$ being the length of the dipole being
described.  The string constraint is imposed by the delta
function, where we use the normalization $\int_{\mathbf{R}^2}\frac{d^2x}{\tau^2}\delta(\frac{x}{\tau})=1$.  The product is over all positive charges in all layers.  
The integration is over the space $\Omega_{\tau}$.  This is defined to be the set of 
all possible spatial configurations of the dipole distribution $\{N_{k,l}\}$ that 
respect the hard-core constraint: no two charges in a given plane may be closer than 
distance $\tau$.

\begin{figure}[ht]
{
\centerline{\includegraphics[angle=0, width=2.5in]{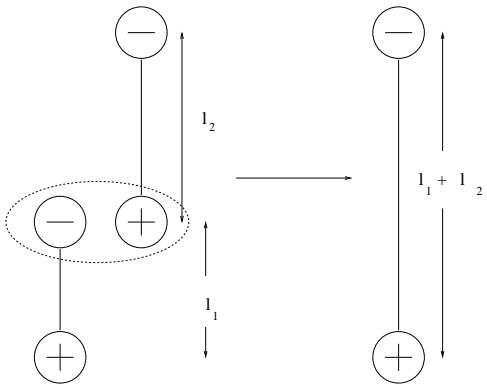}}
\caption{The basic coarse-graining step in our RG transformation.}
\label{fig:rg}}
\end{figure}

Our procedure is an extension of the treatment in
Refs.~\onlinecite{piers},\onlinecite{kt}.  The first part of an RG
procedure normally involves integrating over the high momentum modes
of the system.  In our problem, these correspond to those
configurations where in some plane, we have a pair of charges
separated by a distance between $\tau$ and $\tau+d\tau$.  We assume a
dilute system so only oppositely charged pairs are considered and also
the distance between the members of a pair is taken to be much smaller
than the distance from the pair to another charge.  The basic 
coarse-graining step in our RG transformation is illustrated in 
fig. \ref{fig:rg}.  

Suppose a particular state involves pairing the negative end of an 
$l_1$-dipole in layer $k$ with the positive end of an $l_2$-dipole in layer
$k+l_1$.  Viewed at long length scales, we effectively have an
$(l_1+l_2)$-dipole in layer $k$.  We will find that integrating over
all possible pairings gives a zeroeth order term (which just involves
replacing $\Omega_{\tau}$ with $\Omega_{\tau+d\tau}$) and a number of
correction terms of order $d\tau$ where two short dipoles were
destroyed and replaced by a longer dipole.  Since the procedure
respects the charge neutrality constraint, these correction terms will
combine with other terms in the grand partition sum.  The second step
involves rescaling lengths so that the high momentum cutoff, in the
new variable, is the same as before.  The aim is to see how the
fugacities and couplings change as we run this procedure.

Details of the calculation are given in Appendix B.  Here we give the resulting 
flow equations:
\begin{eqnarray}
\label{flow}
\frac{dy_1}{dt}&=&(2-\kappa)y_1\\ \frac{dy_l}{dt}&=&(2-\kappa)y_l+\sum_{m=1}^{l-1}y_my_{l-m}\\\frac{d\kappa}{dt}&=&0
\end{eqnarray}
where $t=\ln \tau$.  One notable feature is that the coupling does not change with the
flow, in contrast with the 2d Coulomb gas where the coupling does vary (albeit at second
order in the fugacity).  This indicates that strings are stiffer objects than charges.  Another 
observation is that for the initial conditions of our physical problem, namely that $y_1(0)=y_0=2\pi e^{-E_s/k_bT}$ and $y_l(0)=0$ for $l>1$, the flow equations have an exact
solution:
\begin{equation}
y_l=y_0 \tau^{2-\kappa}\Bigl[\Bigl(\frac{y_0}{2-\kappa}\Bigr)(\tau^{2-\kappa}-1)\Bigr]^{l-1}
\label{eq:fug}
\end{equation}
Our RG is valid as long as the corrections to the fugacities are
small, meaning that the derivatives $dy_l/dt$ should be bounded.  If
we look at the above result, Eq.~\ref{eq:fug}, 
we see that when the term in brackets
is greater than 1, $y_l$ diverges with $l$.  Therefore, a critical
length, which we interpret as an in-plane correlation length, is
defined by when the term in brackets equals 1:
\begin{equation}
\frac{y_0}{2-\kappa}(\xi_{\parallel,RG}^{2-\kappa}-1)=1
\end{equation}
Substituting earlier expressions and noting that for our system, $\kappa=1/2$, we
find that:
\begin{eqnarray}
\ln \xi_{\parallel,RG}^2&=&\frac{32g\mu_BJB}{9k_BT}\Bigl(1+\frac{\ln(e^{-E_s/k_BT}+2-\kappa)}{E_s/{k_BT}}\Bigr)\nonumber\\\xi_{\parallel,RG}^2&\sim&\exp(32g\mu_BJB/9k_BT)
\end{eqnarray}
for the fields and temperatures of interest.  This value is the same as that
predicted in Ref.~\onlinecite{msspinice} using a free energy argument.  For
$\tau<\xi_{\parallel,RG}$, $y_l$ decreases with $l$ which means that states
with long dipoles are less probable than states with short dipoles.  If
$\tau>\xi_{\parallel,RG}$, $y_l$ diverges with $l$ which suggests that longer
dipoles are favored, but, as mentioned above, the RG procedure is no longer
valid in this regime.  We note that when $\tau=\xi_{\parallel,RG}$, $y_l$ is
independent of $l$ so that, as in the mean field calculation discussed above,
there is no discernible length scale in the $z$ direction.

If $\tau<\xi_{\parallel,RG}$, then we may consider an out-of-plane length scale, which 
we define nominally as the value of $l=l_{\tau}$ for which $y_l/y_1=1/e$.  
\begin{equation}
l_{\tau}=1+\frac{1}{\ln\Bigl(\frac{\xi_{\parallel,RG}^{3/2}-1}{\tau^{3/2}-1}\Bigr)}
\end{equation}
We may interpret $l_{\tau}$ as the typical length of a string segment that is captured 
by a tube of diameter $\tau$ (where a tube need not be straight).

\subsubsection{Comparison with simulation}

\noindent

\begin{figure}[ht]
{
\centerline{\includegraphics[angle=0, width=3.2in]{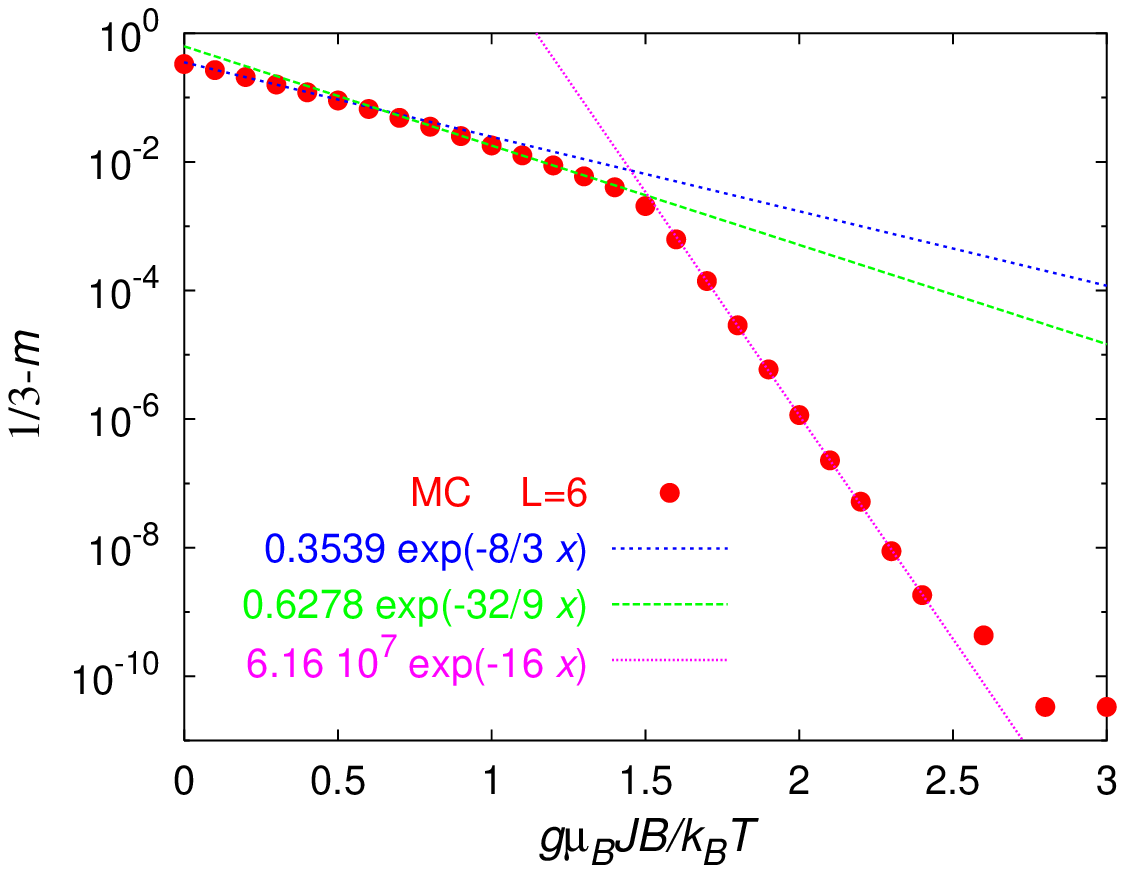}}
\centerline{\includegraphics[angle=0, width=3.2in]{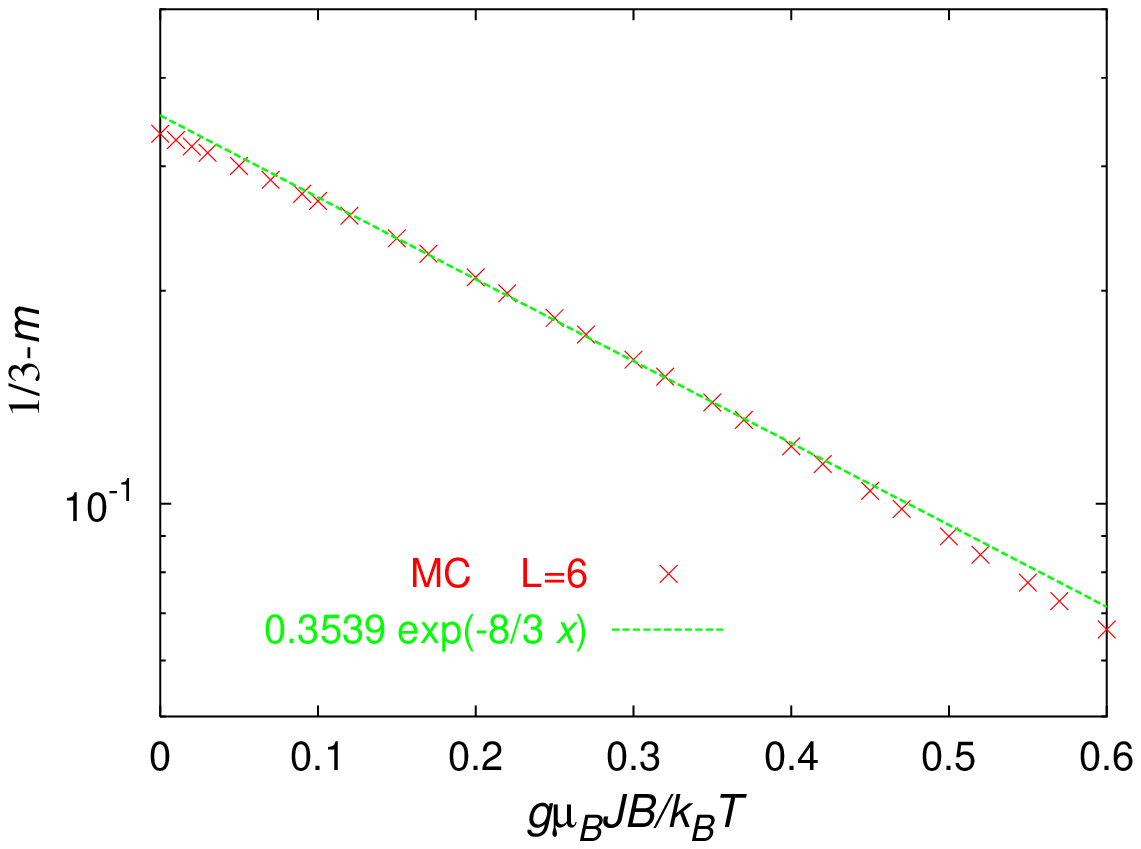}}
\centerline{\includegraphics[angle=0, width=3.2in]{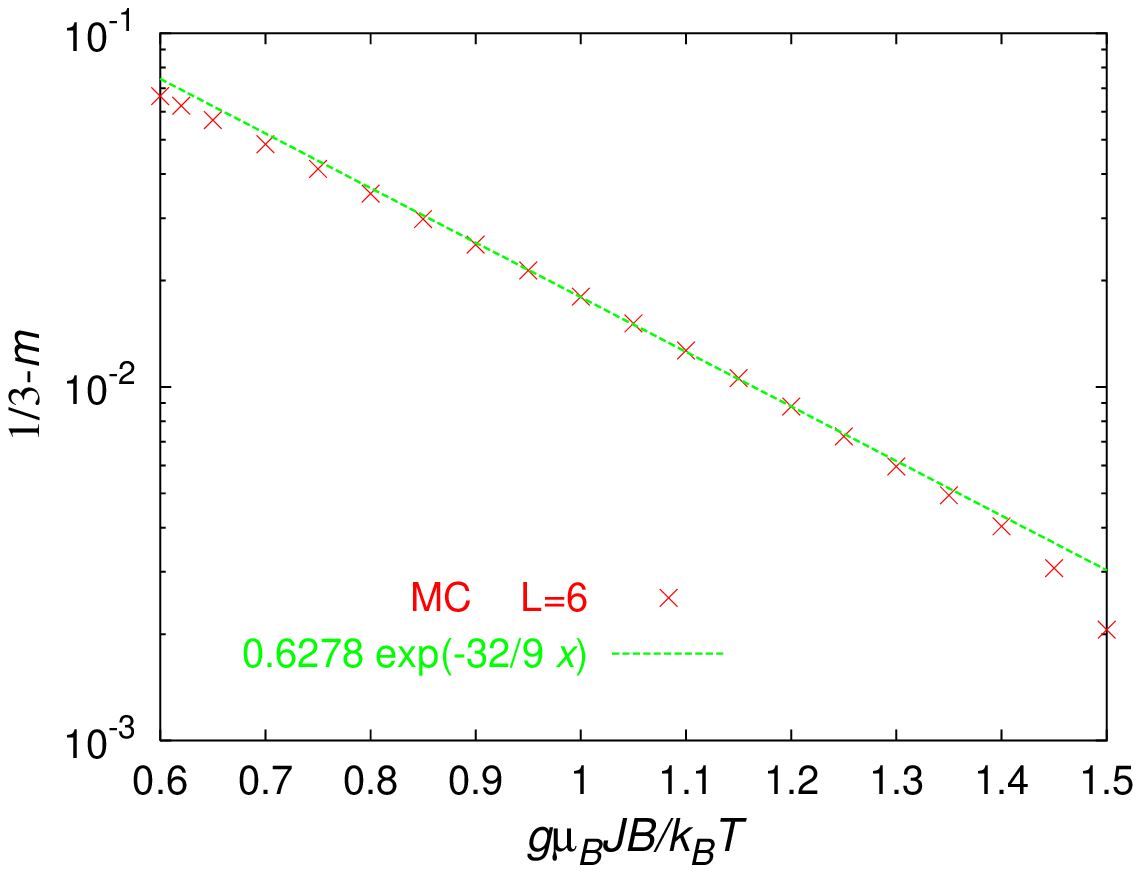}}
\caption{The crossover between exponents.}
\label{fig:crossover}}
\end{figure}

In Fig.~\ref{fig:crossover}, we show the magnetization as a function of
the  magnetic field strength on a log-log scale. Our algorithm allows us
to simulate spin ice in a $[111]$ magnetic field with very high accuracy.

The magnetization should scale with the average density of defects,
which in turn should scale like the inverse square of the in-plane
correlation length.  As shown in this figure, the data at low fields
are well fit by the exponent 8/3 obtained in the mean field
calculation discussed earlier.  At somewhat higher fields, the data
are well fit by the exponent 32/9, obtained by the RG calculation
discussed earlier and also in Ref.~\onlinecite{msspinice} by looking
at the entropic contribution to the free energy.  At high fields, the
exponent of $8L/3$ (=16 for L=6 (sites), as was the case in the simulations)
characterizes a regime where finite-size effects are important, as
discussed below.

The low field crossover makes qualitative sense in that at low fields, there
will be many defects which screen one another which suggests that a mean field
treatment may be reasonably accurate.  At higher fields, the gas of defects is
more dilute so an RG treatment would be required.        

The high field crossover is a finite-size effect since the position of a
crossover between exponents is system size dependent and the corresponding
exponent is also system size dependent, getting steeper with increasing system
size.  The finite-size behavior may be explained as follows. At high magnetic
fields, there are a small number of string defects in the system. The
magnetization and the energy of one string defect in a system of size $L$ are
$-4Lg\mu_{B}J/3$ and $4Lg\mu_{B}JB/3$ respectively. The energy cost grows
linearly with system size and, as mentioned above, the defects are favored
solely due to their entropic contribution to the free energy.  At sufficiently
high magnetic fields, a given system will be too small to provide the entropy
to balance the energy cost of a string.  This will occur when the
magnetization per spin reaches the magnetization of a system with one string
defect:
\bea
 m&=&\left[1/3-2(4L/3)/(16L^{3})\right]g\mu_{B}J \nonumber \\
   &=&\left[1/3-1/(6L^{2})\right]g\mu_{B}J.
\label{mmax}
\eea
In this case, the statistical weight of a single string defect will be a
Boltzmann factor $\exp(-8 L g \mu_{B} J B /3 k_{B} T)$ and the magnetization
will  equal $\left[1/3-C\exp(-8 L g \mu_{B} J B / 3 k_{B}T)\right]g\mu_{B}J$,
where $C$ is some constant. The crossover between different regimes occurs
when the magnetization reaches (\ref{mmax}).  We have good agreement with the
$8L/3$ behavior for a variety of system sizes, including $L=6$ which is shown
in figure ~\ref{fig:crossover}.

\section{The high field regime}
On the plateau, the magnetization of the triangular sublattice is
saturated and we may consider each kagome plane separately.  Thus, the
$3$-dimensional model may be mapped onto a $2$-dimensional
one. Whereas the spins in the triangular sublattice are fixed, the
physics in the kagome planes remains non-trivial. Each triangle on the
kagome plane contains two up pseudospins ($\sigma=1$) and one down
pseudospin ($\sigma=-1$). This Ising model on the kagome lattice may
be mapped onto the dimer model on the hexagonal lattice,\cite{isingquant,
ogata,msspinice} in which a down pseudospin corresponds to a dimer on the
hexagonal lattice. The model retains an extensive ground state
entropy, $S/k_{B}=0.080765$.

If we flip a down (pseudo)spin it violates the ice rule.  This corresponds to 
breaking a dimer into two monomers. 
As with string defects, these monomers may be separated and move freely on the 
lattice.  The energy cost for creating two monomers
is $2{\cal E}=4J_{\rm eff}-2g\mu_{B}JB/3$.  This energy vanishes at a critical
field $B_c=6J_{\rm eff}/(g \mu_{B} J)$. At higher fields the monomers proliferate 
leading to complete saturation and an ordered state with zero entropy. The 
physics near the transition may be described by the following Hamiltonian 
which acts on the kagome lattice:
\beq
  \frac{H}{T}=\sum_{\langle ij \rangle} K_{ij} s_{i} s_{j} - h \sum_{i} s_{i},
\eeq
where the sum is over all nearest neighbors; $s_{i}$ are classical Ising spins
taking values $+1$ and $-1$; $h$ is the strength of a fictitious magnetic
field; and $K_{++}=0$,
$K_{+-}=K_{-+}=K=\left[g\mu_{B}JB/6-J_{\rm eff}\right]/T$, and $K_{--}=\infty$.
The coupling constants imply that each triangle of the kagome lattice contains
at most one down pseudospin and that down spins cost energy (positive or
negative dependent on the magnetic field strength).

\begin{figure}[th]
{
\centerline{\includegraphics[angle=0, width=3.2in]{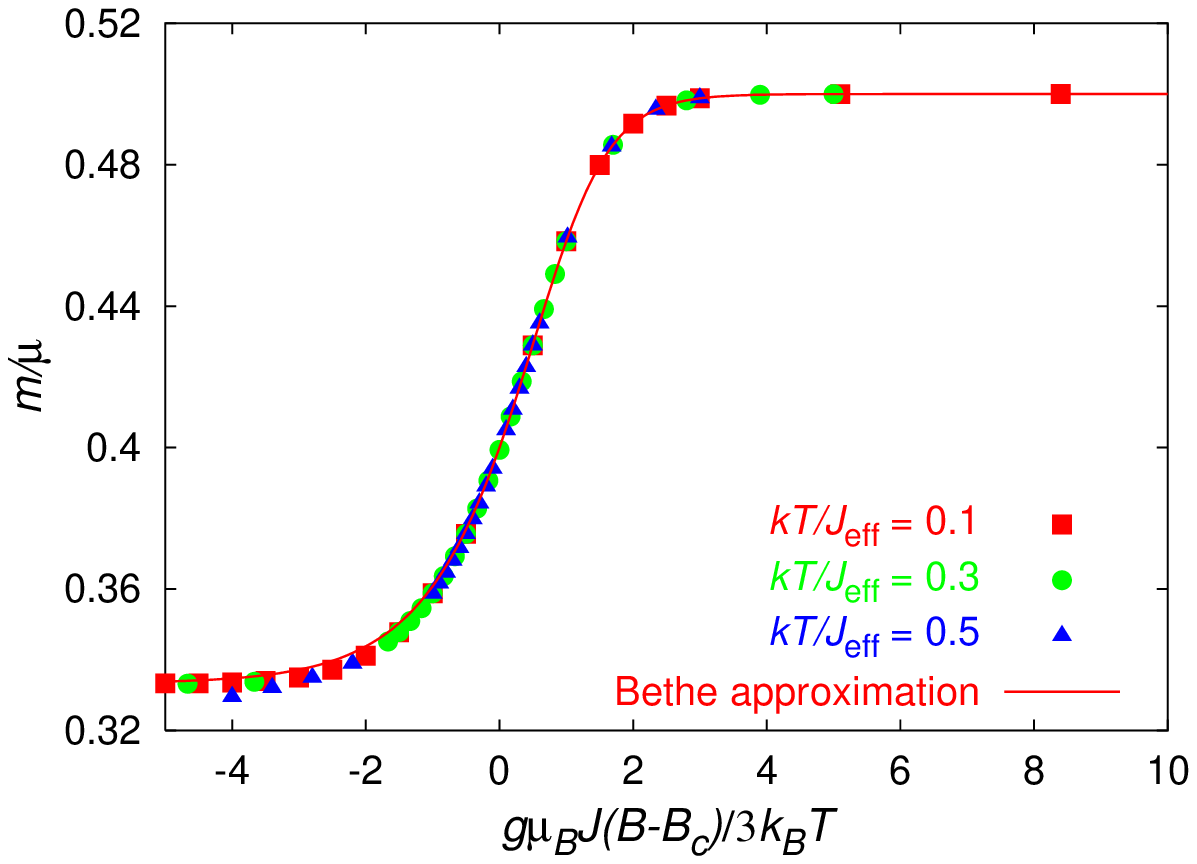}}
\centerline{\includegraphics[angle=0, width=3.2in]{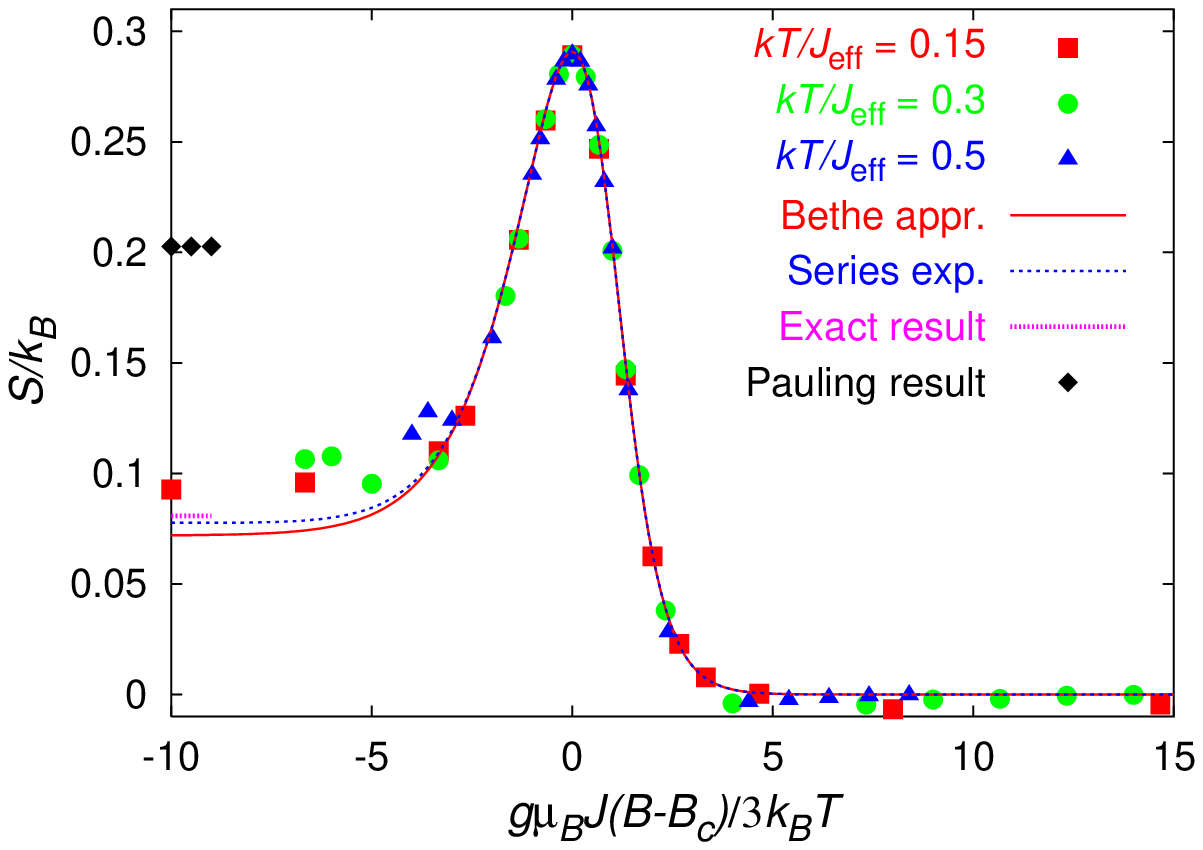}}
\caption{The magnetization (top) and the entropy (bottom) around the
transition between the plateaux. The simple Bethe approximation is
compared to the Monte Carlo results. The exact result for the entropy at
zero monomer density and Pauling's estimate for the entropy at zero magnetic
field are shown for reference. The series expansion contains the results
from Ref.~\onlinecite{naglecorr} on the monomer-dimer model.}
\label{fig:magent}}
\end{figure}

We may calculate the magnetization and entropy using the simple Bethe
approximation.  Details are given in Refs. \onlinecite{naglecorr,sergei}
but we may quote the results:
\bea
   m &=&  \frac{1}{2}\frac{1}{1+x^2}
\label{mag} \\
   S &=&
    -\frac{3 x z \ln{z}}{2+6 x z}+\frac{1}{4}\ln \frac{2z^{3}}{x^{2}(3z-x)}, 
\label{ent}
\eea
where $x=2z/(1+\sqrt{1+8z^{2}})$ and $z=\exp(-2K)$.

In Fig.~\ref{fig:magent}, we compare these expressions with a Monte Carlo
simulation.  The simulation is of a kagome lattice with 16x16 up-triangles
(768 total spins).  The standard single spin-flip Metropolis algorithm was
used, which may explain the inaccuracy in the simulated entropy at low 
fields, where a more clever scheme may be needed to sample the degenerate 
manifold.  The entropy was computed, for a given field, by integrating from 
high temperatures (where $S/k_B=(3/4)\ln 2$ per atom) to low temperatures.   

We find that the simple Bethe approximation is accurate for moderate and high
monomer densities (higher fields) but does not work so well at low monomer
density (lower fields).  As the Bethe approximation does not account for long
cycles on the lattice, the approximation should indeed break down when the
correlation length is large (monomer density is small).  We note that the
correlation length is infinite at zero monomer density since the dimer model
on the hexagonal lattice is critical.

In a higher-order series expansion, one may account for some corrections
to the Bethe approximation.\cite{naglecorr}  As seen in the figure, the 
corrections are almost indiscernible for the magnetization. For the entropy,
the corrections give better agreement at the low monomer density and are 
negligible at high monomer densities.

There is a giant peak in the entropy at the transition point,
$S/k_{B}=1/4\ln(16/5)\approx 0.291$, which exceeds even the zero field
entropy. The peak is due
to the crossing of an extensive number of energy levels which have  
macroscopic entropies.  For $B=B_c$, the 
energies of states corresponding to different numbers of monomer 
defects are equal since the monomer and dimer weights are, by definition, 
equal at the critical field.  There are an extensive number of states 
corresponding to a given number of monomers (below saturation).  The
highly degenerate ground state manifold explains the large spike in the 
entropy.

\section{Crossing points}

The theory described in the previous section implies that the curves of
magnetisation versus field, plotted for different temperatures, will display a
crossing point.  This arises simply because the partition function depends on
magnetic field and temperature effectively only through the combination
$(B-B_c)/T$. Thus, when plotted as a function of $B-B_c$, the curves coincide
only at the point $B=B_c$. At this point, the Bethe approximation gives a
value for the magnetisation of $m=0.4 g\mu_B J$, see Eq.~\ref{mag}.

In addition, we expect a crossing point at low fields, due the interplay of
string and monomer defects. Indeed, where the plateau is well-formed, the
string density is $n_{s}\sim\exp(-32g\mu_{B}JB/9k_{b}T)$ and the monomer
density is $n_{m}\sim \exp(-8{\cal E}_{m}/7k_{B}T)$, where ${\cal
  E}=g\mu_{B}J(B_c-B)/3$ is the energy of creating one monomer. The crossing
point occurs when $n_{s}=n_{b}$. With logarithmic accuracy, we can write 
\beq
\frac{32g\mu_{B}JB}{9k_{b}T} = \frac{8g\mu_{B}J(B-B_c)}{21k_{B}T}.  
\eeq Thus
the crossing point lies at $B^{\star}=3B_c/31$.

\section{Relation to experiment and other theories}

Our model gives a description of the high field transition that is
qualitatively consistent with experiment for a range of temperatures
\cite{hiroiice}. In particular, a peak in the entropy has been
observed close to the high-field termination of the plateau (Fig. 9 in
Ref.~\onlinecite{hiroiice}).  As this feature was taken to be an
experimental artefact, it was not analyzed in detail in that
work. However, it appears that its height is rather smaller than the
one we find here, although the number of data points is not enough
to determine the center of the peak or its height.

However, recent experiments\cite{sakakiice} on the spin ice
compound Dy$_2$Ti$_2$O$_7$ have indicated that at low temperatures, the high
field transition becomes first order.  In Ref.~\onlinecite{sakakiice}, the
onset of first order behavior was found to occur for temperatures lower than a
critical temperature of $T_c\sim 0.36$K ($\sim 0.327 J_{eff,Dy}/k_B$).  Figure
\ref{fig:magent} shows that our predicted curves remain continuous even at
temperatures below this observed $T_c$.

A likely reason for the discrepancy is the long range nature of the dipolar
interaction, which we approximated as a nearest neighbor Ising model.  The
simplest way to account for these interactions is to model the ignored
interaction terms as giving rise to a magnetic field proportional to the
magnetization.  By assuming the magnetization $M$, as a function of the
effective field $B+\alpha M$, has the same functional form as given in figure
\ref{fig:magent}, we may self-consistently determine $M$ for a given $B$.
Using $\alpha$ as a free parameter, we find that this simple model predicts
the onset of first order behavior, at the experimentally observed critical
field $B_c$, only for temperatures in the millikelvin range.  To obtain a
higher numerical $T_c$ requires a larger $\alpha$, which causes a lower
numerical $B_c$.  To get the numerical $T_c$ to match experiment requires an
$\alpha$ so large that our numerical $B_c$ is ``negative'' (in the sense of
artificially extending the $M=1/3$ line of figure \ref{fig:magent} for the
purpose of a spline fit).  It seems that a more careful treatment of the
dipolar interaction is required in order to explain the recent experimental
results. Also, we have not considered the impact of the slowdown of the
dynamics which is observed at low temperature.\cite{dynamicspapers}

As for the crossing points mentioned above, the high-field one does indeed
appear to be present in the experimental data\cite{matsuice,sakakiice} in the
appropriate temperature range.  The experimental value of the magnetization at
the crossing point is about $m=0.38 g\mu_B J$, reasonably close to the
theoretical value $m=0.4 g\mu_B J$. By contrast, a crossing point at small
fields is harder to make out, and an approximate estimate of its location
gives $B^{\star}=0.35B_c$, in disagreement with the theoretical
$B^{\star}=3B_c/31$.

\section{Entropy spike and magnetocalorics}
Fig.~\ref{fig:magent} shows a stark contrast between the behavior of
magnetization and entropy as the field strength is increased. Whereas the
magnetization increases monotonically going from one plateau to the other, the
entropy displays a strong (but smooth) non-monotonicity. 

One question which naturally arises is whether such an entropy peak exists
more generally between two magnetization plateaus -- what is the crucial
ingredient for the existence of the spike? The sectors with different
magnetizations are degenerate because not only do the monomer defects not cost
any energy at the degeneracy point, but they also do not interact. Such a
situation has in fact been observed already in a much more familiar frustrated
model, namely the triangular Ising antiferromagnet in a longitudinal field.
Here, there is a (non-degenerate) low-field plateau with magnetization of 1/3,
in addition to the usual saturated high-field plateau. These two are separated
by a degeneracy point where `up-up-up' and `up-up-down' triangles are
degenerate.\cite{metcalf}
The statistical mechanics of that point is described by the
hard-hexagon model,\cite{hardhex} the entropy of which is extensive. A similar
phenomenon -- a magnetization plateau bounded by two entropy spikes -- also
appears in the case of an effectively 1d
helimagnet.\cite{diptimanheli} 

In classical Ising models, such a degeneracy seems not so surprising
as the allowed energies are naturally discrete. However, a similar
situation can arise even in bona-fide Heisenberg models. This follows
from the result by Richter\etal,\cite{richteretal} who demonstrated
that near saturation, on a range of frustrated lattices (including the
kagome), localized spin-1/2 excitations exist. As one sweeps the
magnetic field from saturation downwards, one would therefore also
expect an entropy spike in those models. A numerical study testing
this assertion is in progress.\cite{derzhkorichter}

\subsection*{Cooling by adiabatic (de)magnetization}

At low temperatures, near the degeneracy point, the partition function depends
on magnetic field and temperature effectively only through the combination
$(B-B_c)/T$. One may thus argue that the spike may be used to effect cooling
by adiabatic demagnetization\cite{zhitodemag} 
in exactly the same way one may use paramagnets
-- analogous constraints limit the application in either case.

There are two features which may be worth pointing out at this point.
Both follow from the fact that -- unlike in the case of a paramagnet
-- $B_c\neq 0$.  Firstly, maximal cooling occurs at a finite field,
namely around $B_c$.  This phenomenon may therefore be useful to
effect cooling for a magnet in a field, with the restriction that
$B_c$, for a given spin ice compound, is not tunable.  
Secondly, if $B$ approaches $B_c$ from below, one
can in fact obtain ``cooling by adiabatic magnetization'', as entropy
and magnetization grow together in this regime.

\section{Conclusions}
In this paper, we have analyzed in detail the magnetization curve of
nearest-neighbor spin ice in a [111] magnetic field. The basic ingredient
which makes this system particularly interesting is that a uniform
field can be used to couple to the Ising pseudospins as a staggered
field.\cite{chacha,icerm} This amounts to the possibility of applying fields
which would have appeared to be rather unnatural in the formulation of a
simple Ising model (without the detour via spin ice) on the pyrochlore
lattice.

As a result, one observes an attractively rich behavior. Perhaps the most
salient is the dimensional reduction from pyrochlore to kagome under the
application of an external field.  The restoration of three-dimensionality
upon weakening the field goes along with the one-dimensional string defects.
We hope that the extension developed here of Kosterlitz's RG treatment to such
extended defects might be of more general use.

A particularly attractive feature of the monomer-dimer model we have obtained
here lies in the fact that the relative monomer and dimer fugacities in the
low-temperature ($T\ll J_{\rm eff}$) regime are given by simple Boltzman
weights of Zeeman energies. They are thus straightforwardly tunable by
changing the strength of the applied field. In particular, anisotropic
fugacities can be obtained by tilting the field, and they therefore do not
require an actual manipulation (such as an application of anisotropic stress)
of the two-dimensional layer.

As discussed previously in Ref. \cite{msspinice} the price for our ability
to analyze the model in such detail has been the omission of the long-range
nature of the dipolar interaction. A truncation of the interaction at only the
nearest-neighbor distance would seem a rather drastic step; an expectation of
quantitative agreement between experiment and the nearest-neighbor model will
in general likely be misplaced. However, as we argue in a different context,
it turns out that, in an intermediate temperature regime, this is not entirely
unreasonable.\cite{imslargen} This observation might lie at the basis of the
fact that the measured dipolar ice entropy agrees so well with Pauling's
estimate. Our `prediction' of the entropy peak between the intermediate and
saturated plateaux bears witness to the promise of our approach to unearth 
at least some qualitative features of interest.

\section*{Acknowledgements}
We would like to thank Michel Gingras, Hans Hansson, Ryuji Higashinaka, Peter
Holdsworth, Johannes Richter, Anders Karlhede, Satoru Nakatsuji for useful
discussion.  This work was in part supported by the Minist\`ere de la
Recherche et des Nouvelles Technologies with an ACI grant.  SLS would like to
acknowledge support by the NSF (DMR-0213706) and the David and
Lucile Packard Foundation.

\appendix
\section{The cluster algorithm}
We use a loop algorithm to simulate spin ice at low fields. The algorithm
probes only the spin ice ground state manifold and therefore can work only at
low temperatures and low magnetic fields. All attempted loop flips are
accepted in our algorithm.

The algorithm works as follows. To construct a loop, we, first, pick at
random a tetrahedron of fixed orientation (and mark it as a first
tetrahedron in a loop), then we pick with probability $1/2$ a spin direction
(in or out of a tetrahedron) and pick a first spin in a loop using the
following rules. If both spins with the chosen direction are on the kagome
sublattice then we pick the spin with a probability $1/2$, which is
independent of the spin orientation. If one spin is on the triangular
sublattice and another is on the kagome sublattice then we pick the spin
with probability that depends on the spin orientation. Namely, if the spin
on the triangular sublattice is out of the tetrahedron (along the magnetic
field), we pick the spin on the kagome or triangular sublattices with
respective probabilities
\beq
  p_{1} = \frac{1}{1+g},
\label{pkag}
\eeq
and
\beq
  p_{2} = \frac{g}{1+g}
\label{ptri}
\eeq
where $g$ will be fixed by the detailed balance condition, see below, and
$p_{1}+p_{2}=1$. If the spin on the triangular sublattice points into
the tetrahedron, we pick the spin on the kagome or triangular sublattices
with probabilities $p_{2}$ and $p_1$ respectively. Then we flip the chosen
spin thus introducing two defects in the tetrahedra that share the spin.

After choosing the first spin, we move to the neighboring tetrahedron with a
defect. The next tetrahedron has two spins with the opposite orientation. We
flip one of these two spins adding it to the loop using the same prescription
as we used to pick the first spin. Thus we move the defect to another
tetrahedron. Then we repeat this procedure iteratively moving one of the two
defects through the lattice until we encounter the other defect in the first
tetrahedron -- the two defects will annihilate and the loop will be closed.
Since we add spins to the loop with alternating signs -- two spins with
opposite orientation from each tetrahedron we traverse, the ice rule is not
violated.

The algorithm is ergodic since any pair of different configurations differ
by spins on closed loops only. They can always be connected by flipping these
loops.

\begin{figure}[ht]
{\vspace{0.5cm}
\centerline{\includegraphics[angle=0, width=2.4in]{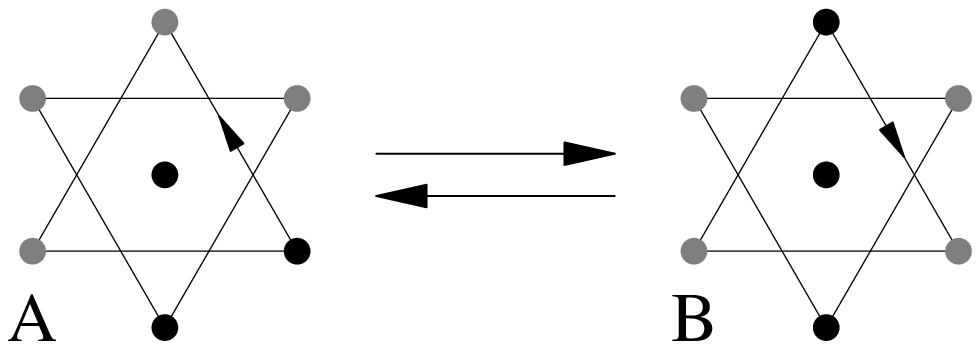}}
\caption{Configurations $A$ and $B$. Tetrahedra are shown on top of each
other. Small arrows indicate a short sequence of a loop. Up and down spins
are denoted by black and grey dots.}
\label{fig:detbal1}}
\end{figure}

\begin{figure}[ht]
{\vspace{0.5cm}
\centerline{\includegraphics[angle=0, width=2.4in]{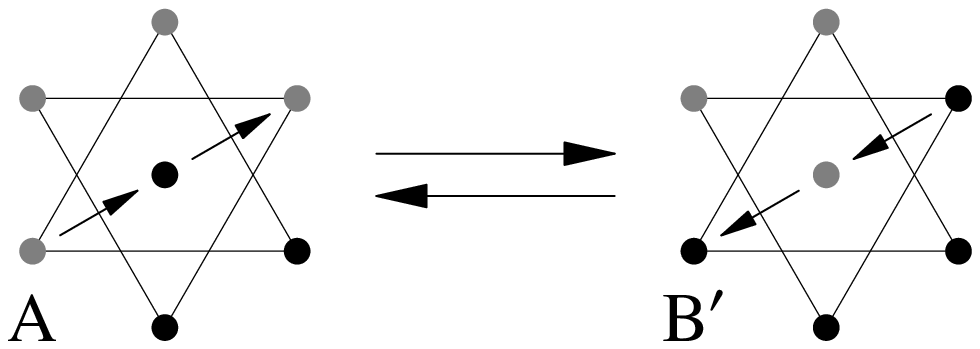}}
\caption{Configurations $A$ and $B'$. Tetrahedra are shown on top of each
other. Small arrows indicate a short sequence of a loop. Up and down spins
are denoted by black and grey dots.}
\label{fig:detbal2}}
\end{figure}

Let us sketch the proof of the detailed balance condition. Suppose that we
have flipped some loop. In order to prove detailed balance, the
first site in a loop that returns us to the original configuration must be
the first site in the original loop and the reversed loop must be constructed
in the reverse direction. We can prove the detailed balance condition locally,
i.e. for all short sequences of the loop, see Fig.~\ref{fig:detbal1} and
Fig.~\ref{fig:detbal2}. It is easy to check that most of these sequences are
trivial, i.e. they have equal energies before and after spin flip and equal
probabilities to
go from one to another configuration. An example of such a simple sequence is
shown in Fig.~\ref{fig:detbal1}. The probability of going from configuration
$A$ to configuration $B$ is equal to the probability of going from $B$ to $A$
(equal to $1/2$). In order to prove the detailed balance condition, we only
need to consider the energies of single spins that are the second spins in the
sequences (the energies of the first spins in the sequences are taken into
account in the previous step). These spins have the same
energies. Thus the detailed balance condition is satisfied trivially.
An example of a nontrivial sequence is shown in Fig.~\ref{fig:detbal2}. The
energies of configurations $A$ and $B'$ are different there.
We have to prove the detailed balance condition
\beq
  P(A\rightarrow B)/P(A\leftarrow B')=P(B')/P(A)
\label{detbal}
\eeq
The right hand side in (\ref{detbal}) is just a ratio of Boltzmann weights
and is equal to $\exp(8h/3)$, where $h=g\mu_{B}J B/k_{B}T$, since the energy
of configuration $A$ (the energy of the second and third spins in the
sequence) is $4hk_{B}T/3$, and the energy of configuration $B'$ (the energy of
the second and third spins in the sequence)
is $-4hk_{B}T/3$. According to our algorithm, the probability of going from
configuration $A$ to configuration $B'$ is $P(A\rightarrow B')=p_{1}/2$ and the
reverse probability of going from $B'$ to $A$ is $P(A\leftarrow B')=p_{2}/2$.
We have from (\ref{detbal})
\beq
  g=\frac{p_{2}}{p_{1}}=\exp(-8h/3).
\eeq
Therefore if we choose $p_{1}$ and $p_{2}$ as
\beq
  p_{1} = \frac{1}{1+e^{-8h/3}},
\eeq
and
\beq
  p_{2} = \frac{e^{-8h/3}}{1+e^{-8h/3}}
\eeq
then the detailed balance condition is fulfilled.

\section{RG Calculation}
We introduce the abbreviation:
\begin{equation}
d\Omega_{\tau}=\prod_{k,i\in I_k}\Biggl(\frac{d^2x_{k,i}^{(1)}}{\tau^2}\frac{d^2x_{k,i}^{(2)}}{\tau^2}\delta\Bigl(\frac{x_{k,i}^{(1)}-x_{k,i}^{(2)}}{\tau}\Bigr)\Biggr)
\end{equation}
in terms of which the canonical partition function for a given dipolar 
distribution $\{N_{k,l}\}$ may be written: $Z(\{N_{k,l}\},\tau)=\int_{\Omega_{\tau}}d\Omega_{\tau}\exp(-H)$.  Our RG calculation has two steps.  The first step is integrating over short length scales, i.e. those states where at least one pair of charges is
separated by a distance between $\tau$ and $\tau+d\tau$.  The second step is to rescale
variables to restore the short distance cutoff.  When we carry out the first step, the result is a zeroth
order term and a correction of order $d\tau$:
\begin{equation}
Z(\{N_{k,l}\},\tau)=\int_{\Omega_{\tau+d\tau}}d\Omega_{\tau}\exp(-H)+\sum_{k,l,m,i,j}I_{klmij}
\end{equation}
where $I_{klmij}$ is the contribution of the configuration that has the negative end 
of the $i$th $m$-dipole of layer $k$ paired with the positive end of the $j$th $(l-m)$-dipole
of layer $k+m$.  The sum over $k$ is over all planes; the sum over $l$ is over all dipole lengths
up to the number of planes; and the sum over $m$ is from 1 to $l-1$.  The form of this term is given by:
\begin{eqnarray}
\label{klmij}
&I_{klmij}&=\int_{\Omega_{\tau+d\tau}^{'}}d\Omega_{\tau}^{'}e^{-H'}\int_{A}\frac{d^2x_i^{(2)}}{\tau^2}\delta\Bigl(\frac{x_i^{(1)}-x_i^{(2)}}{\tau}\Bigr)\nonumber\\&\times&\int_{d(x_i^{(2)},\tau)} \frac{d^2x_j^{(1)}}{\tau^2}\delta\Bigl(\frac{x_j^{(1)}-x_j^{(2)}}{\tau}\Bigr)e^{-H(x_i^{(2)},x_j^{(1)})}\nonumber\\
\end{eqnarray}
The region of integration of the positive charge $x_j^{(1)}$ is an annulus of radius $\tau$ 
and thickness 
$d\tau$ centered on the negative charge $x_i^{(2)}$.  This region is denoted by $d(x_i^{(2)},\tau)$.  The
position of this negative charge (and hence the pair) is integrated over the entire area $A$.  Strictly
speaking, $x_i^{(2)}$ would have to avoid the hard cores of all of the other charges but this introduces
an error of order $(d\tau)^2$.  $\Omega_{\tau+d\tau}^{'}$ is the space of configurations of the rest
of the charges in which the charges are separated from each other by a distance of at least 
$\tau+d\tau$.  $H(x_i^{(2)},x_j^{(1)})$ refers to the piece of the Hamiltonian which involves 
charges $x_i^{(2)}$ and $x_j^{(1)}$ and the rest of the Hamiltonian is denoted by $H'$.

The $x_j^{(1)}$ integration amounts to making the substitution $\vec{x}_{j}^{(1)}=\vec{x}_{i}^{(2)}+\vec{\tau}$; $d^2x_{j}^{(1)}=\tau d\tau d\theta$; and integrating over angles.  If we denote the latter
two of integrals of equation \ref{klmij} by $I$, then:
\begin{eqnarray}
I&=&\frac{d\tau}{\tau}\int_{A}\frac{d^2x_i^{(2)}}{\tau^2}\int_0^{2\pi}d\theta e^{-H(\vec{x}_i^{(2)},\vec{x}_i^{(2)}+\vec{\tau})}\nonumber\\ &\times&\delta\Bigl(\frac{x_i^{(1)}-x_i^{(2)}}{\tau}\Bigr) \delta\Bigl(\frac{\vec{x}_i^{(2)}-\vec{x}_j^{(2)}+\vec{\tau}}{\tau}\Bigr)
\end{eqnarray}

We assume that our gas of defects is sufficiently dilute that the following distances are much greater
than the pair separation $\tau$: (1) the distance of a particle in plane $k+m$ from our pair, (2) the
distance of a particle in plane $k$ from the positive charge $x_i^{(1)}$, and (3) the distance of 
a particle in plane $k+l$ from the negative charge $x_j^{(2)}$.  In this dilute limit, we may make 
the approximation:
\begin{eqnarray}
\delta\Bigl(\frac{\vec{x}_i^{(1)}-\vec{x}_i^{(2)}}{\tau}\Bigr)\delta\Bigl(\frac{\vec{x}_i^{(2)}-\vec{x}_j^{(2)}+\vec{\tau}}{\tau}\Bigr)\approx \frac{\tau^2}{A}\delta\Bigl(\frac{\vec{x}_i^{(1)}-\vec{x}_j^{(2)}}{\tau}\Bigr)\nonumber\\
\end{eqnarray}
We also have that $H(x_i^{(2)},x_j^{(1)})$ is small in this limit, which allows us to expand the
exponential and to leading order, the integral may be done exactly \cite{piers}.  The result is:   
\begin{eqnarray}
I&=& \frac{d\tau}{\tau}\delta\Bigl(\frac{x_i^{(1)}-x_j^{(2)}}{\tau}\Bigr)\Bigl(2\pi-\frac{(\pi\kappa\tau^2)^2}{A}\sum_{a\neq b} e_ae_b\ln\frac{r_{ab}}{\tau}\Bigr)\nonumber\\&\approx&2\pi \frac{d\tau}{\tau}\delta\Bigl(\frac{x_i^{(1)}-x_j^{(2)}}{\tau}\Bigr)\nonumber\\
\end{eqnarray}
In the penultimate line, the sum refers to a sum over all charges, positive
and negative, residing in the plane $k+m$.  This sum term may be neglected in the large
$A$ limit, which is why, in contrast to the Kosterlitz calculation\cite{kt}, the coupling
strength does not vary during our RG flow (see equation \ref{flow}).  The delta function 
implies that the $m$-dipole and $(l-m)$-dipole have been combined into a larger $l$-dipole.
Returning to our correction term:
\begin{eqnarray}
I_{klmij}\approx 2\pi\frac{d\tau}{\tau}\Biggl[\int_{\Omega_{\tau+d\tau}^{k,l,m}}d\Omega^{k,l,m}_{\tau}\exp(-H)\Biggr]\nonumber\\
\end{eqnarray}
where the space $\Omega_{\tau+d\tau}^{k,l,m}$ is analogous to $\Omega_{\tau+d\tau}$, except that there
is one less $m$-dipole in layer $k$; one less $(l-m)$-dipole in layer $k+m$; and one more $l$-dipole
in layer $k$.  What we are actually interested in is the grand partition function (equation \ref{grand}).
Because our RG procedure is consistent with the charge neutrality constraint, the various $\{I_{klmij}\}$
may be combined with different terms in the grand partition function.  When we substitute into 
equation \ref{grand} and arrange terms, we find that:
\begin{eqnarray}
&\mathcal{Z}&=\sum_{\{N_{\kappa,\lambda}\}}\frac{1}{\prod_{\kappa,\lambda}(N_{\kappa,\lambda})!}\Bigl[\int_{\Omega_{\tau+d\tau}}d\Omega_{\tau}\exp(-H)\Bigr]\nonumber\\&\times&\Biggl[\prod_{\kappa,\lambda}\Bigl(\frac{y_{\lambda}}{2\pi}\Bigr)^{N_{,\lambda}}\nonumber\\&+&\sum_{k,l,m}\Bigl[\prod^{'}_{\kappa,\lambda}\Bigl(\frac{y_{\lambda}}{2\pi}\Bigr)^{N_{,\lambda}}\Bigr]2\pi\frac{d\tau}{\tau}\frac{y_my_{l-m}}{(2\pi)^2}N_{k,l}\Bigl(\frac{y_{l}}{2\pi}\Bigr)^{N_{k,l}-1}\Biggr]\nonumber\\
\end{eqnarray} 
The prime on the second product means that $y_l^{N_{k,l}-1}$ has been taken outside the product.  If
the fugacities are small, then we may write this in a more convenient way:
\begin{eqnarray}
\mathcal{Z}&=&\sum_{\{N_{k,l}\}}\Bigl[\prod_{k,l}\frac{(y_l+\frac{d\tau}{\tau}\sum_{m=1}^{l-1}y_my_{l-m})^{N_{,l}}}{(2\pi)^{N_{k,l}}(N_{k,l})!}\Bigr]\nonumber\\&\times&\int_{\Omega_{\tau+d\tau}}d\Omega_{\tau}\exp(-H)
\end{eqnarray}
Finally, we rescale lengths, $x\rightarrow x(1+d\tau/\tau)^{-1}$, and find 
(dropping primes):
\begin{eqnarray}
\mathcal{Z}=\sum_{\{N_{k,l}\}}\Bigl[\prod_{k,l}\frac{\Bigl(\frac{y_l^{'}}{2\pi}\Bigr)^{N_{,l}}}{(N_{k,l})!}\Bigr]\int_{\Omega_{\tau}}d\Omega_{\tau}\exp(-H)
\end{eqnarray}
where
\begin{eqnarray}
y_l^{'}=(y_l+\frac{d\tau}{\tau}\sum_{m=1}^{l-1}y_my_{l-m})(1+2\frac{d\tau}{\tau})(1-\kappa\frac{d\tau}{\tau})
\end{eqnarray}
The flow equations (\ref{flow}) follow from this.


\begin{thebibliography}{99}
\bibitem{coarse}
We neglect the term proportional to $\cos(2\pi h)$.
\end{thebibliography}
\end{document}